\documentclass[aps,prc,twocolumn,showpacs,nofootinbib,floatfix,superscriptaddress]{revtex4-1}
\usepackage{graphicx}
\usepackage{amsfonts}
\usepackage{amsmath}
\usepackage{hyperref}
\usepackage{float}
\usepackage{verbatim} 

\begin{document}

\title{Mach cones in viscous heavy-ion collisions}

\author{I. Bouras}
\affiliation{Institut f\"ur Theoretische Physik,
Johann Wolfgang Goethe-Universit\"at,
Max-von-Laue-Str.\ 1, 60438 Frankfurt am Main, Germany}

\author{B. Betz}
\affiliation{Institut f\"ur Theoretische Physik,
Johann Wolfgang Goethe-Universit\"at,
Max-von-Laue-Str.\ 1, 60438 Frankfurt am Main, Germany}

\author{Z. Xu}
\affiliation{Department of Physics, Tsinghua University, Beijing 100084, China}
\affiliation{Collaborative Innovation Center of Quantum Matter, Beijing 100871, China}

\author{C. Greiner}
\affiliation{Institut f\"ur Theoretische Physik,
Johann Wolfgang Goethe-Universit\"at,
Max-von-Laue-Str.\ 1, 60438 Frankfurt am Main, Germany}

\begin{abstract}

The formation of Mach cones
is studied in a full $(3+1)$-dimensional setup of ultrarelativistic
heavy-ion collisions, considering a transverse and longitudinal
expanding medium at Relativistic Heavy-Ion Collider energies. 
For smooth initial conditions and central collisions
the jet-medium interaction is investigated using high-energy
jets and various values of the ratio of shear
viscosity over entropy density, $\eta/s$.
For small viscosities, the formation of Mach cones
is proven, whereas for larger viscosities the characteristic structures
smear out and vanish eventually. The formation of a double-peak
structure both in a single- and in a multiple-jet event is discussed.

\end{abstract}


\date{\today}

\maketitle

\section{Introduction}
\label{sec:introduction}

Collisions of heavy nuclei both at the Relativistic Heavy-Ion Collider (RHIC) \cite{Adams:2005dq,Arsene:2004fa,Adcox:2004mh,Back:2004je}
and the Large Hadron Collider (LHC) \cite{PhysRevLett.105.252301,PhysRevLett.105.252302,PhysRevLett.106.032301,PhysRevLett.107.032301}
indicate the formation of a new state of strongly interacting matter, 
the quark-gluon plasma (QGP). The large value of the measured elliptic flow coefficient,
$v_2$, suggests that the QGP behaves like an almost
perfect fluid as demonstrated by various calculations of viscous hydrodynamics \cite{Luzum:2008cw,Song:2008hj,Niemi:2011ix,Schenke:2011tv} and microscopic
transport models \cite{Xu:2007jv,Xu:2008av,Xu:2011fi}. The values determined for the ratio of
shear viscosity over entropy density are
approximately $\eta/s = 0.1-0.2$ and thus close
to the conjectured lower bound, $\eta/s = 1/4\pi$,
derived via the correspondence between conformal field theory and string theory
in an anti-de Sitter (AdS) space \cite{Kovtun:2004de}.

High-energy jets traversing the expanding medium deposit energy and momentum.
Because of this interaction with the bulk medium, those jets are (strongly) suppressed,
a phenomenon that is called jet quenching \cite{Adams:2003kv,Adams:2005dq,Arsene:2004fa,Adcox:2004mh,Back:2004je,Bjorken,Gyulassy:1990dk,Wang:1991xy,Wang:1991vs}. 
The explicit mechanisms of the jet-medium interaction are still a matter of research. The two- and three-particle
correlations extracted from experiment show a complete suppression of the away-side jets for $3<p_T<6$~GeV
and due to energy and momentum conservation the signal appears again
in a lower-$p_T$ range, showing a double-peak structure
\cite{Wang:2013qca,Wang:2004kfa,Adams:2005ph,Adler:2005ee,Ulery:2005cc,Ajitanand:2006is}. The origin of the double-peak structure was discussed to be 
connected to Mach cones generated by fast partons traversing the
strongly coupled medium 
\cite{Stoecker:2004qu,Scheid:1974zz,Hofmann:1976dy,Baumgardt:1975qv,Gutbrod:1989gh,Neufeld:2011yh,Neufeld:2010tz,Noronha:2008un,Ma:2006fm,Bouras:2010nt,Ruppert:2005uz,Koch:2005sx,Neufeld:2011yh,Renk:2005si,Gubser:2007ga,CasalderreySolana:2004qm,Zhang:2007qx,Li:2009ax,Rau:2010qy,Betz:2010qh,CasalderreySolana:2007km}. However, recent studies on the triangular flow and hot spots
\cite{Wang:2013qca,Ma:2010dv,Schenke:2010rr,Bhalerao:2011bp,Takahashi:2009na,Andrade:2009em,Alver:2010gr,Gyulassy:1996br}
issued a further, probably more satisfactory explanation, for the appearance of the double-peak structure.

Up until today, however, the contribution of jet-induced Mach cones
to the double-peak structure remains unclear. To gain a better understanding
of the contribution of Mach cones to the double-peak structure in two-particle 
correlations, we investigate the evolution pattern of jet-induced Mach cones
and their particle distribution in a relativistic, ($3+1$)-dimensional expanding system of a 
heavy-ion collision for various values of the ratio of shear viscosity over entropy
density, $\eta/s$, using the microscopic transport model BAMPS 
(Boltzmann approach of multiparton scatterings) \cite{Xu:2004mz}.

In an earlier publication \cite{Bouras:2012mh}, we already studied the formation
of Mach cones using the above-mentioned kinetic transport model applying
a static box scenario and thus neglecting all effects from expansion.  
We showed that, considering a source term depositing both energy and momentum,  
the double-peak structure is overshadowed by the strong contribution of the head shock 
and diffusion wake, while the double-peak structure appears for 
a source term (that lacks physical motivation from any theoretical model)
which leads to only energy but no momentum deposition. 
These results are consistent with an investigation based on an AdS/Conformal Field Theory calculation
\cite{Noronha:2008un} and a hydrodynamic model \cite{Betz:2008wy,Betz:2008ka}. The 
comparison with the latter publications proves once more that BAMPS can model a hydrodynamic expansion 
\cite{Bouras:2010hm,Bouras:2009nn}. Using the same non-expanding scenario, we studied the effect of dissipation
in a box scenario \cite{Bouras:2012mh} and found that 
dissipation tends to destroy any Mach cone signal.

However, as discussed in Ref.~\cite{Satarov:2005mv},
the flow-velocity profile created by jets in the transverse plane
interacts with the radial flow of the background medium, changing,
e.g., the effective angle of a Mach cone.
This effect of transverse expansion on
Mach cones has also been studied in
Refs.~\cite{Betz:2010qh}. It was found that a double-peak
structure can be created by averaging over different jet paths. 
The main contribution to the double-peak structure originates from
jets that are deflected by radial flow. In the rare case that
a jet traverses into the opposite direction of the radial flow, the 
interplay between radial flow and jets reduces the strong contribution
of the diffusion wake and head shock and results in a double-peak structure
for this single event. 

In the following, we investigate a similar setup as in Ref.~\cite{Betz:2010qh},
but including not only transverse but also longitudinal expansion, exploring 
in particular the influence of viscosity. 
In this work, the units are $\hbar = c = k = 1$.

\section{Numerical setup}
\label{sec:numercialSetup}

We use the framework of BAMPS in a
fully ($3+1$)-dimensional setup designed for
ultrarelativistic heavy-ion collisions \cite{Xu:2004mz}.
Here, we treat a massless gluon gas as classical Boltzmann
particles with a degeneracy factor of $g = 16$.
Particles only collide via binary collisions with an
isotropic cross section, i.e., 
a cross section with an isotropic distribution of the
collision angle. We perform
the numerical calculations using a constant value
of the ratio of shear viscosity over entropy density, $\eta/s$.
For isotropic binary collisions, the shear viscosity
is given by $\eta = 0.4\,e \, /(n\sigma)$ \cite{deGroot_book},
where $e$ ($n$) is the local rest frame energy (particle) density.

As we intend to focus on the impact of transverse radial and longitudinal flow on jets, 
we neglect additional effects from density fluctuations. Likewise, we neglect 
any effects originating from elliptic flow and only consider
central collisions. We apply smooth Glauber initial conditions in the transverse direction
\cite{Miller:2007ri}
and a Gaussian rapidity distribution in the longitudinal direction as it was found at RHIC
\cite{Bearden:2004yx} that the rapidity distribution of charged hadrons can be described
by a Gaussian. In the beam ($z$) direction, we determine the width of the nuclear overlap
region based on the Lorentz-contracted nuclear thickness that is approximated 
by a Gaussian as well. (Please note that we initialize the system
with a finite width. The system, however, will behave boost-invariantly near
$z=0$ at times $<1$~fm.)
For the high-$p_T$ region we apply a power law which approximately
fits the $p+p$ data \cite{Adler:2003pb,Adams:2006xb}. Thus, we apply the following
parametrization for the initial
non-thermal single-particle distribution function 
\begin{equation}
\label{eq:parametristationHIC}
\begin{split}
f(\vec{x},\vec{p}) =& K \frac{1}{E} \left ( \frac{Q^n}{Q^n + p_T^n} \right )^m \exp{\left(- \frac{y_{\rm rap}^2}{\sigma_y^2} \right)}  \\
 & \times \exp{\left(- \frac{z^2}{\sigma_z^2} \right)} T_{\rm A}\left(x + \frac{b}{2}, y \right) T_{\rm B}\left(x - \frac{b}{2} , y\right) \, ,
\end{split}  
\end{equation}
where $p_T = \sqrt{p_x^2 + p_y^2}$ denotes the transverse momentum,
$y_{\rm rap}$ the momentum rapidity, and $b$ the impact parameter. 
The nuclear thickness function \cite{Kolb:2001qz} 
\begin{equation}
T_{\rm A}\left(x,y\right) = \int_{-\infty}^{+\infty} dz \rho_{\rm A}(x,y,z) \,
\end{equation}
is integrated in the $z$-direction. $\rho_{\rm A}(\vec{x})$ denotes the Woods-Saxon 
distribution for the nucleus $A$,
\begin{equation}
\rho_{\rm A}(\vec{x} ) = \frac{\rho_0}{1 + \exp{\left( \frac{\left|\vec{x} \right| - R_{\rm A}}{D} \right)}} \,.
\end{equation}
Here, $R_{\rm A} = 1.12 A^{1/3} - 0.86 A^{-1/3}$
and the mean density of the nucleus is $\rho_0 = 0.17 \, \rm fm^{-3}$.
For the thickness parameter we use $D = 0.54$ fm. As we are studying RHIC energies, 
we exclusively use gold nuclei with a mass number of $A = B = 197$
and choose 
$Q=1.3 \, \rm GeV$, $n=4$, $m=1.5$, $\sigma_y=1$, $\sigma_z=0.13 \, \rm fm $, and
$K = 0.0135$. 

In the following we investigate scenarios where the jet is
set on top of the bulk medium.
Please note that the jet interacts
with the same cross section as the medium particles.
%
\begin{figure}[tp!]
\begin{center}
\includegraphics[width=\columnwidth]{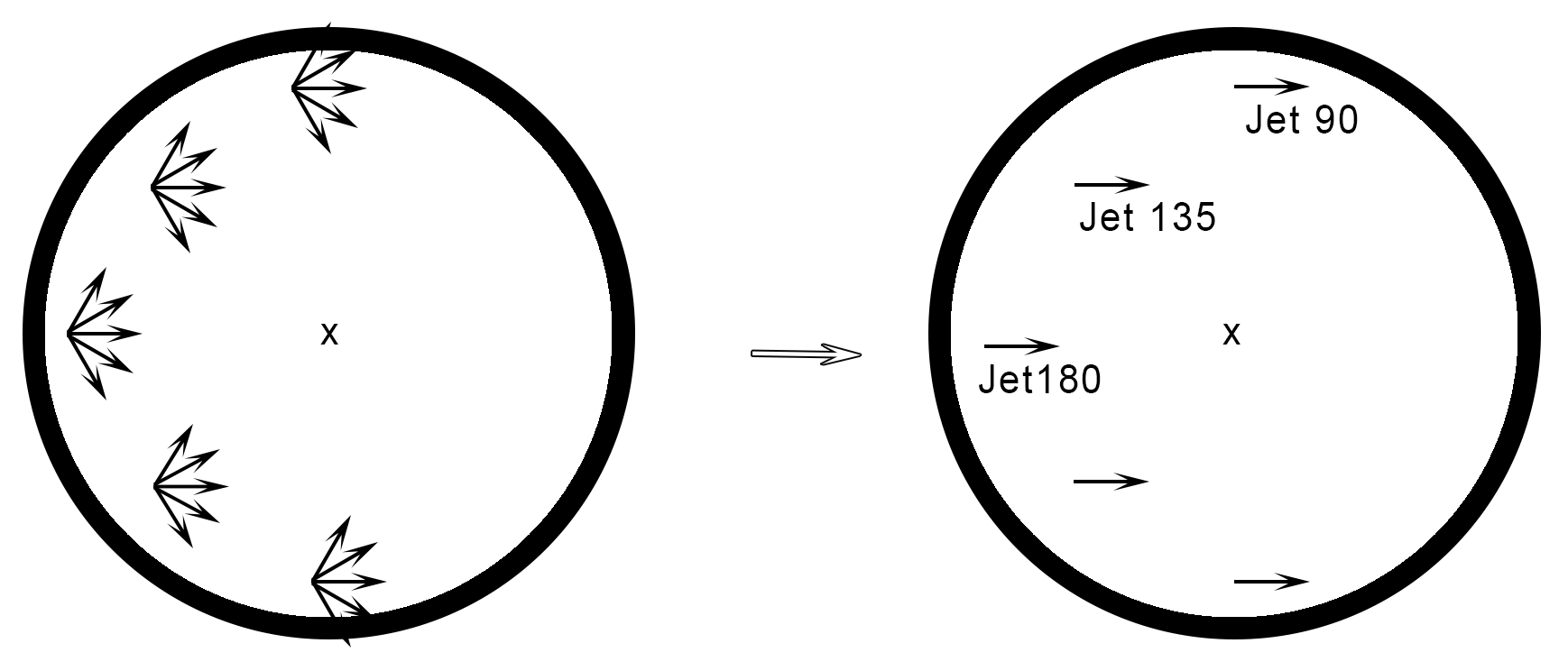}
\caption[Schematic representation of different jet paths]
{(Color online) Schematic representation of different jet paths initialized on a semi-circle;
see also Ref.\ \cite{Betz:2010qh}.
The left-hand panel shows all possible paths, and the right-hand panel shows the
reduced number of paths due to reasons of symmetry.}
\label{fig:numMach:hic_MachCone_InitSemiCircle}
\end{center}
\end{figure}
%

Similar to the numerical setup discussed in Ref.~\cite{Betz:2010qh} we
initialize the starting points of the jet on a semi-circle
displayed in Fig.~\ref{fig:numMach:hic_MachCone_InitSemiCircle} with
a radius of $r = 4$ fm at mid-rapidity,
\begin{equation}
\vec{x}_{\rm jet} = r
\begin{pmatrix}
\cos{\phi_{\rm jet}}\\ \sin{\phi_{\rm jet}} \\ 0
\end{pmatrix}
\textbf{.}
\end{equation}
While in the experiment back-to-back correlated jets are created due to momentum
conservation, we assume that the near-side jet escapes rapidly to the vacuum 
and thus, we neglect the near-side jet contribution. In contrast, the
away-side jet traverses the hot and dense medium of the collision.

Because of symmetry, the possible jet paths that need to be studied 
in a central collision (see left-hand panel of Fig.~\ref{fig:numMach:hic_MachCone_InitSemiCircle}) reduces drastically (as shown in the right-hand panel of Fig.~\ref{fig:numMach:hic_MachCone_InitSemiCircle}).
All jets studied here have an initial momentum only in the $x$-direction of 
$p_x = E_{\rm jet} = 20$ GeV.

Although performing a full ($3+1$)-dimensional simulation,
we restrict the following discussion on the midrapidity region. Thus,
we extract all quantities and azimuthal particle distributions for
a small space-time rapidity of $\left| \eta_{\rm rap} \right| < 0.1$.
Since the parton cascade BAMPS has no effective hadronization
process implemented yet, the final particle distribution is obtained
by stopping the simulation at a certain time and extracting the
macroscopic quantities as well as the 
azimuthal particle distribution from 
the final gluon-momentum distribution.

\section{Results}

In the following we study the evolution of jet-induced Mach cones in
ultrarelativistic heavy-ion collisions for three different scenarios in
order to gain a better understanding of the contribution of 
transverse and longitudinal expansion on the evolution of Mach cones
induced by high-energy jets, considering various values of the ratio of
shear viscosity over entropy density, $\eta/s$.

\subsection{Scenario I}

\begin{figure*}[tp!]
\begin{center}
\includegraphics[width=\textwidth]{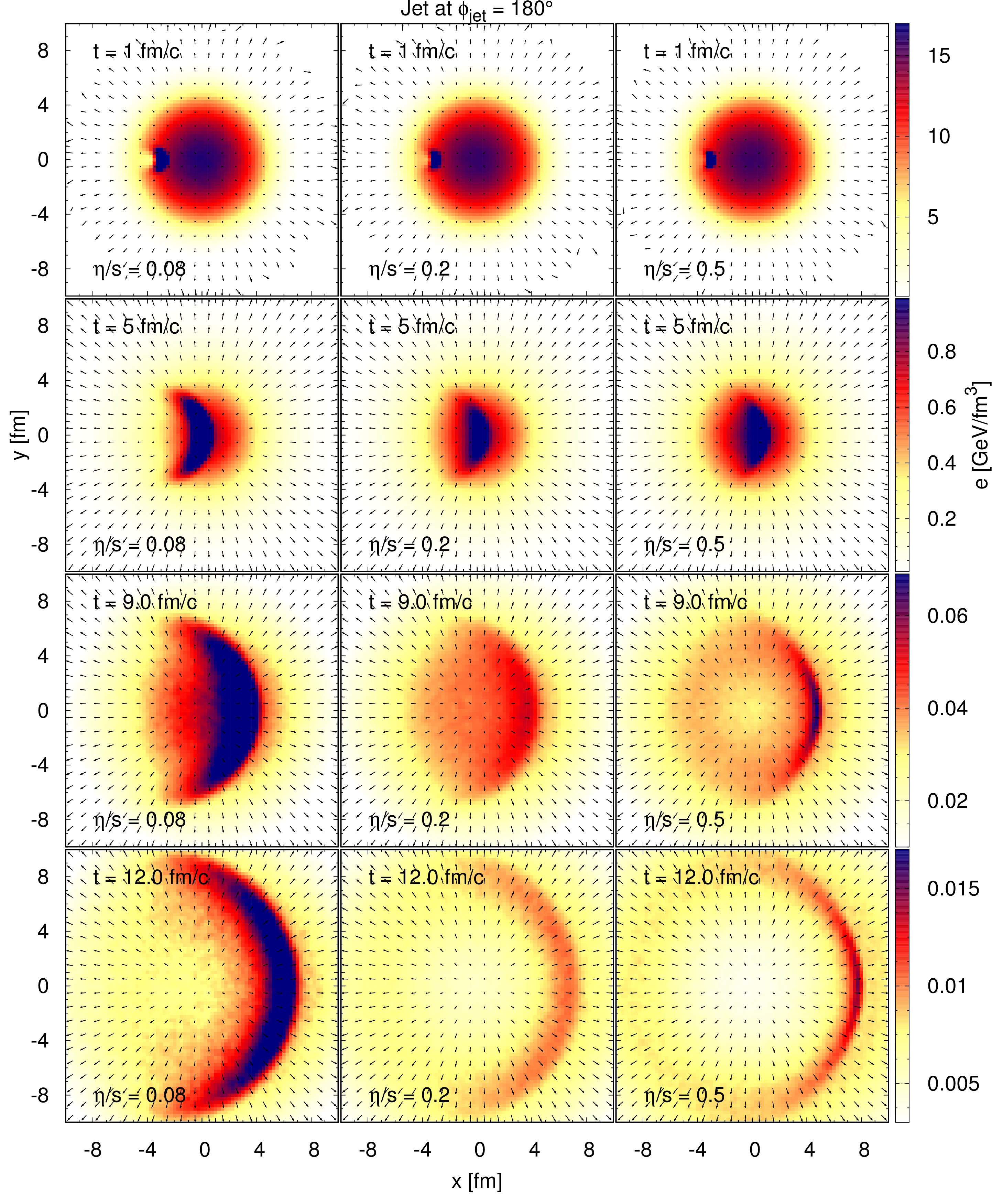}
\caption[Time evolution of Mach cones in central HIC with a jet
starting at a fixed angle position of
$\phi_{\rm jet} = 180^{\circ}$ on the semi-circle]
{(Color online) Time evolution of a Mach cone in a central heavy-ion collision. 
The local rest frame energy density around midrapidity 
($\left| \eta_{\rm rap} \right| < 0.1$) is shown, overlaid by the 
velocity profile, indicated by scaled arrows.
The results are depicted at various time steps
and different values of the ratio of shear viscosity
over entropy density, $\eta/s$.
The jet is initialized at a fixed-angle position of $\phi_{\rm jet} = 180^{\circ}$ on the semi-circle with 
an initial momentum of $p_x = E_{\rm jet} = 20$ GeV.}
\label{fig:numMach:hic_jetB_evolution_Fixed180}
\end{center}
\end{figure*}

In a first scenario we consider a jet starting at a
fixed-angle position of $\phi_{\rm jet} = 180^{\circ}$
on a semi-circle. In this particular case the jet initially propagates
in the opposite direction to the radial flow generated in a heavy-ion collision. We examine
different values of the ratio of shear viscosity over entropy density,
reflecting different possible interaction strengths with the medium. To
illustrate the results we depict a time evolution of the local-rest frame energy-density
profile around mid-rapidity, $\left| \eta_{\rm rap} \right| < 0.1$,
in Fig.~\ref{fig:numMach:hic_jetB_evolution_Fixed180}, overlaid with
a velocity profile represented by scaled arrows.

At early times, $t \sim 1$ fm/$c$, the energy
density reaches very high values and a Mach cone has not yet built
up. The energy behind the jet, however, is significantly lowered
for $\eta/s = 0.08$, indicating that matter is pushed in the forward direction. 
This effect is reduced with increasing values of $\eta/s$.

%
\begin{figure}[ht!]
\begin{center}
\includegraphics[width=\columnwidth]{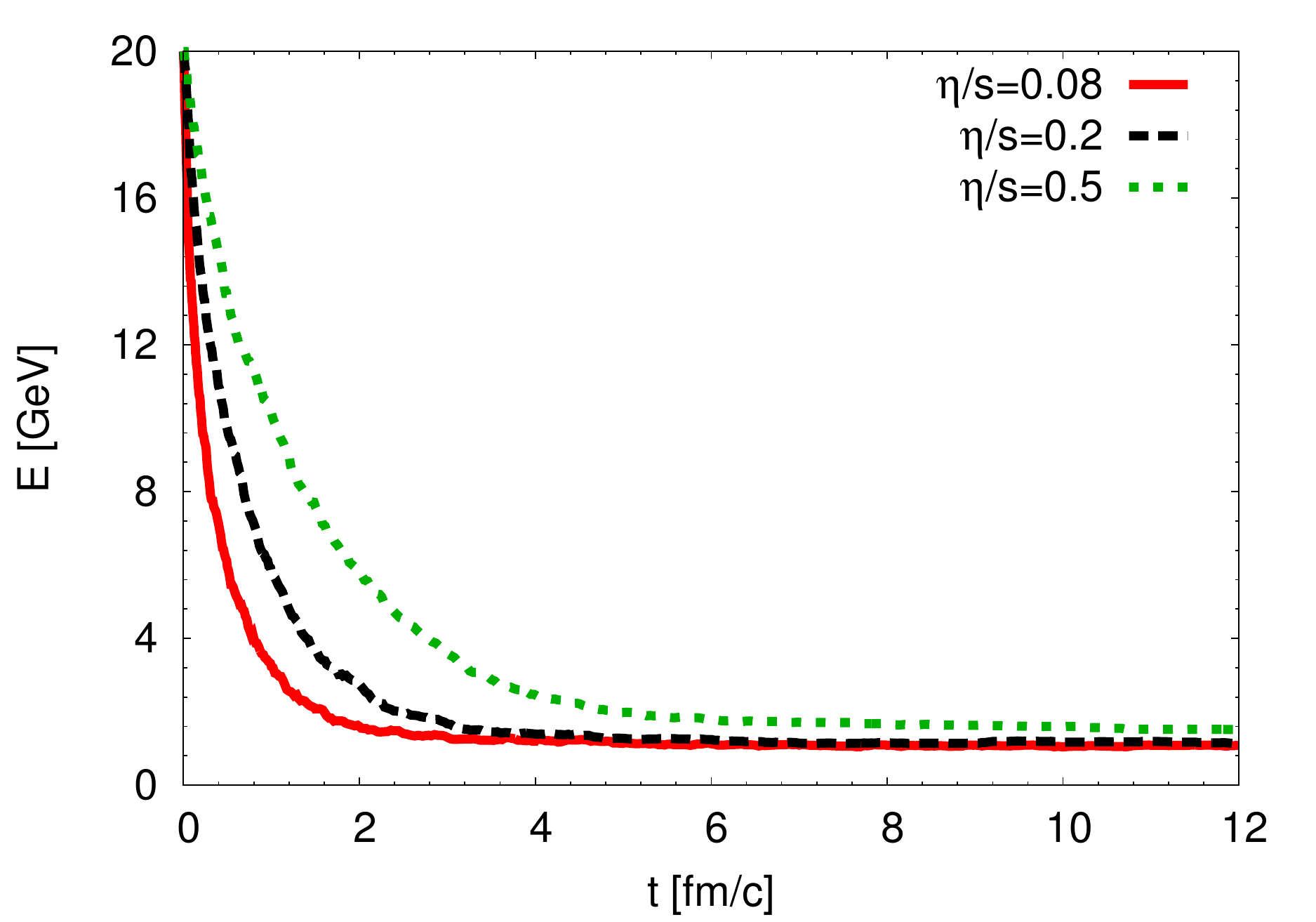}
\caption{(Color online) The time evolution of the jet energy extracted from BAMPS calculations
shown in Fig.~\ref{fig:numMach:hic_jetB_evolution_Fixed180} for different
values of the ratio of shear viscosity over entropy density, $\eta/s$.}
\label{fig:numMach:hic_jetB_eLoss_Fixed180}
\end{center}
\end{figure}
%

At later times, the energy density decreases drastically 
due to the longitudinal and the transverse expansion.
A conical structure induced by the jet has developed at $t = 5 $~fm/$c$
for $\eta/s = 0.08$. In contrast, for larger values of $\eta/s$,
such a structure has not built up (yet). The differences in the
shape of the Mach cone for various values of $\eta/s$ becomes more evident
at later times, $t = 9$ and $t=12$~fm/$c$. In the case of $\eta/s = 0.08$, the energy
density in the region of the developed shock front is
much more increased than for larger values of $\eta/s$. 
Interestingly enough, the maximal energy density reached in the head-shock
region for $\eta/s = 0.2$ is smaller than for $\eta/s = 0.5$. The reason 
is that the interaction of the jet with the medium is weakened for $\eta/s = 0.5$;
i.e., the jet is almost not quenched. In turn this means that the 
jet propagates faster for larger values of $\eta/s$.
For the case with $\eta/s  = 0.08$ the initial jet has already lost almost
all of its energy, i.e., has thermalized, within a very short time in
the beginning of the simulation. This effect is demonstrated
in Fig. \ref{fig:numMach:hic_jetB_eLoss_Fixed180}, where 
the corresponding time evolution of the jet energy for different values
of $\eta/s$ is shown. As mentioned above we do not distinguish
between bulk medium and jet particles, which implies that for $\eta/s  = 0.08$
the mean free path of the jet is very small. On the other side,
the larger the $\eta/s$, the smaller is the energy loss of the jet.
It is clear that
considering perturbative QCD based elastic and inelastic
processes \cite{Fochler:2013epa,Fochler:2010wn} would
change significantly the jet energy loss. However, such
a treatment remains a future task.

%
\begin{figure}[ht!]
\begin{center}
\includegraphics[width=\columnwidth]{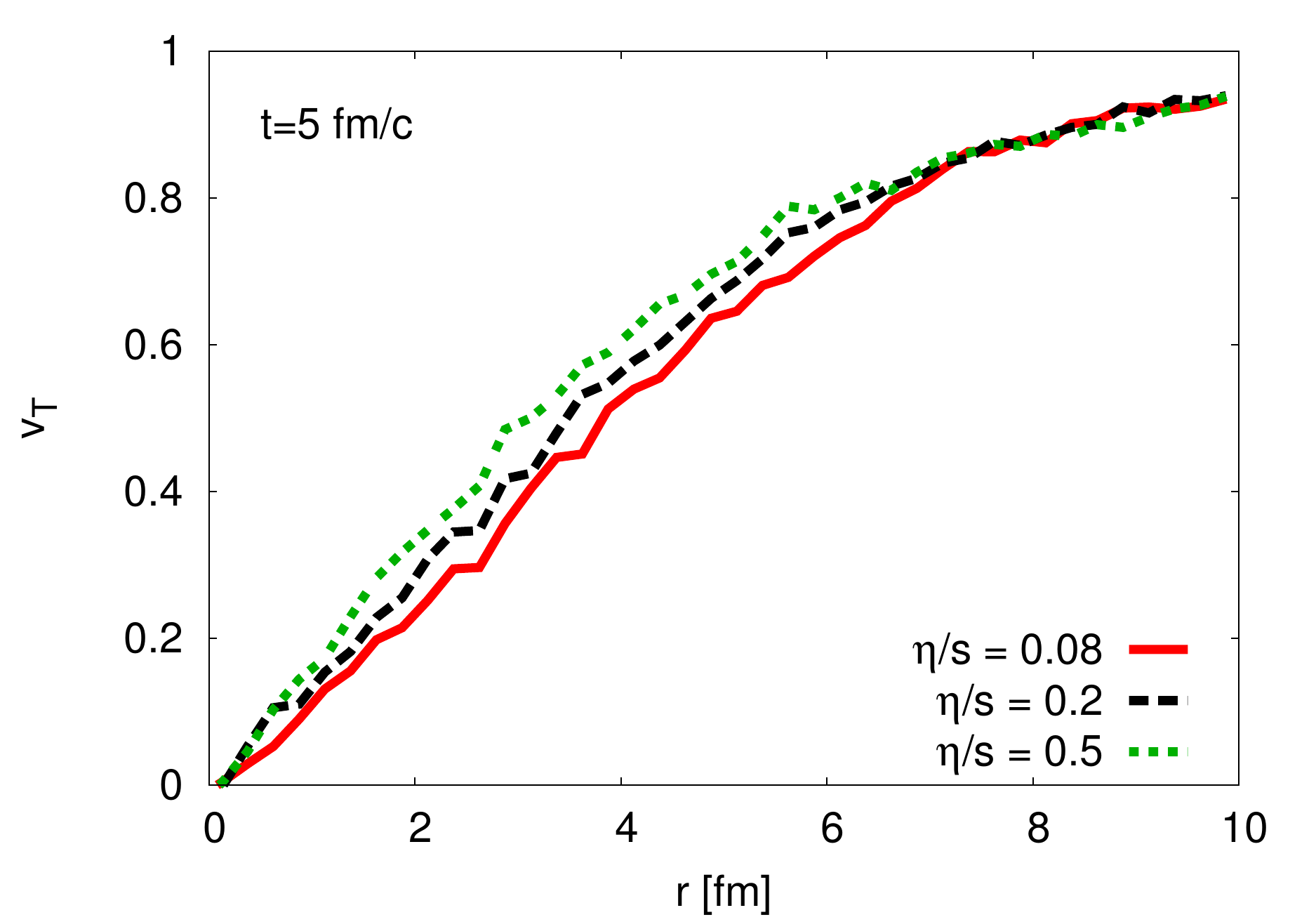}
\caption{(Color online) The extracted radial flow from BAMPS calculations at $t = 5$ fm /$c$
and central collisions for different
values of the ratio of shear viscosity over entropy density, $\eta/s$.}
\label{fig:numMach:hic_radialFlow}
\end{center}
\end{figure}
%

The results in Fig.~\ref{fig:numMach:hic_jetB_evolution_Fixed180}
indicate that a single jet propagating through the middle of the medium
creates a Mach cone for moderate values of
$\eta/s$ and sufficiently large times.
The shock front, however, is strongly curved due to
the interaction of the jet with the medium. The diffusion wake
is not directly visible as it is superimposed by the radial flow of the background medium, 
indicating that the contribution of the diffusion wake and head shock 
is possibly reduced in the final particle distribution. After passing the center
of the collision, however, the radial flow broadens the jet-induced shock front region.
Please note that there are two competing effects here. As shown in
Fig.~\ref{fig:numMach:hic_radialFlow} at $t = 5$ fm /$c$, the radial flow
increases with larger $\eta/s$, while the interaction of
the jet with the background medium decreases for larger $\eta/s$. Thus,
the strength of the radial flow changes between the different $\eta/s$
scenarios studied here.

%
\begin{figure}[ht!]
\begin{center}
\includegraphics[width=\columnwidth]{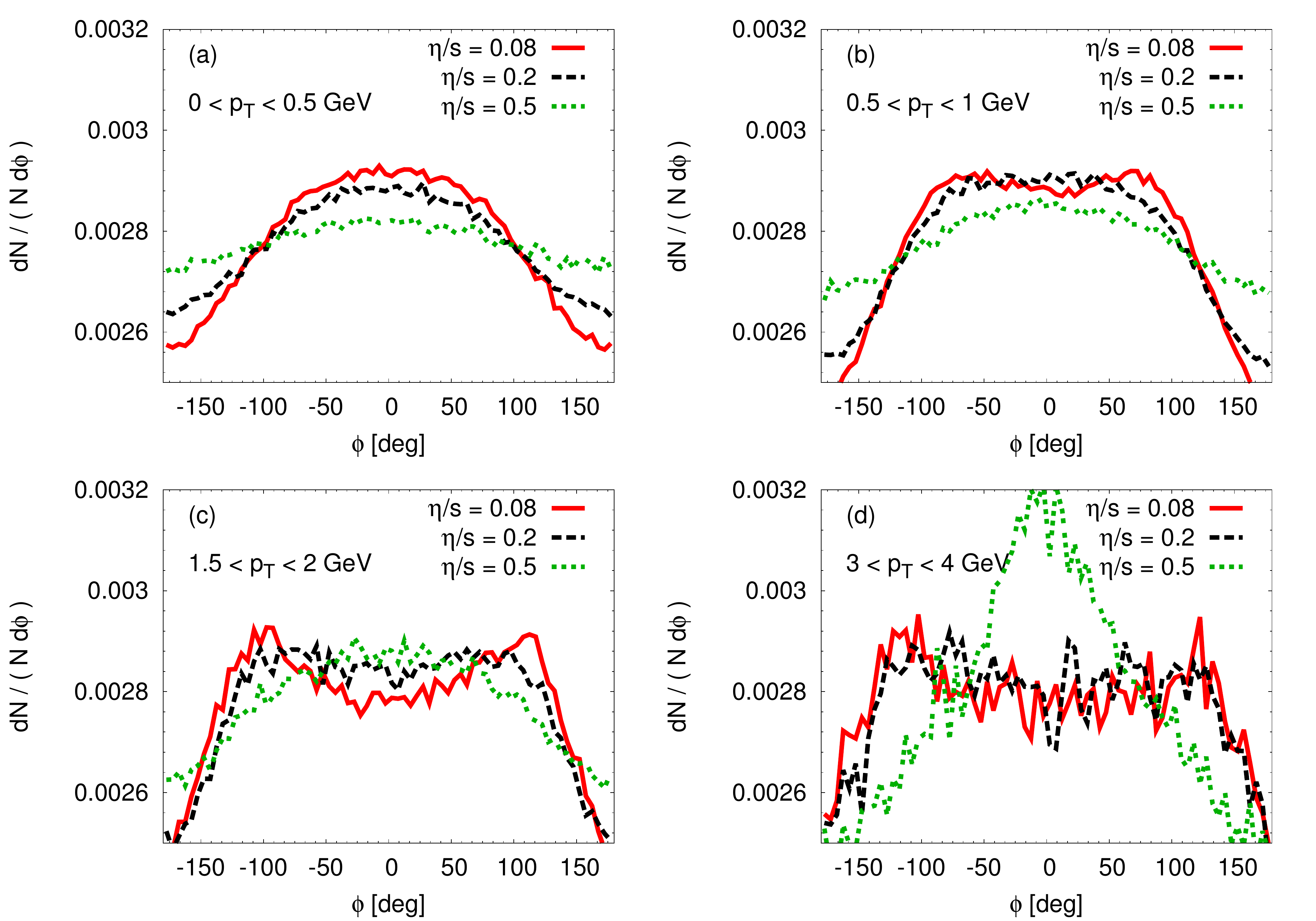}
\caption[Two-particle correlations for several $p_T$ regions extracted
from BAMPS calculations for jets in HIC for a fixed angle position
of $\phi_{\rm jet} = 180^{\circ}$ on the semi-circle]
{(Color online) Two-particle correlations, ${\rm d}N/(N{\rm d}\phi)$, extracted from BAMPS calculations
for different values of $\eta/s$ and $p_T$-cuts.
The jet is initialized at a fixed-angle position of $\phi_{\rm jet} = 180^{\circ}$ on the semi-circle (see Fig.~\ref{fig:numMach:hic_jetB_evolution_Fixed180}).
The results are shown for a fixed time $t=12$ fm/$c$.}
\label{fig:numMach:hic_TPC_jetB_eCuts_Fixed180_5GeV}
\end{center}
\end{figure}
%

In Fig.~\ref{fig:numMach:hic_TPC_jetB_eCuts_Fixed180_5GeV}, we show the extracted 
normalized azimuthal particle distribution ${\rm d}N/(N{\rm d}\phi)$.
The plot demonstrates that a double-peak structure can develop for a small
value of the ratio of shear viscosity over entropy density, $\eta/s$,
and a $p_T$-cut of $p_T>0.5$~GeV.
For $\eta/s = 0.08$ and $0.5 < p_T < 1$ GeV
a double-peak structure is observed at $\phi \approx \pm 70^\circ$,
suggesting that the contribution of head shock and diffusion wake is indeed compensated by
the radial flow. For larger transverse momenta, however [see, e.g.,
Fig.~\ref{fig:numMach:hic_TPC_jetB_eCuts_Fixed180_5GeV} (d)],
the double-peak structure appears around $\phi \approx \pm 120^\circ$.
Here, this conical structure is due to the fact that matter flows into a
region of lower pressure and energy density behind the jet and is thus not
directly a signal of the Mach cone. Finally, for $\eta/s = 0.5$ 
the contribution from the head shock and diffusion wake is very strong, leading
to a single peak in the forward direction.

The above discussion reveals that a Mach cone can only develop if the
energy of the jet is neither too high nor too low. In the former case, the 
strong contribution of the head shock and diffusion wake leads to
a single peak in the jet direction, whereas in the latter case the radial flow
of the background medium distorts a Mach cone signal.

\begin{figure}[ht!]
\begin{center}
\includegraphics[width=\columnwidth]{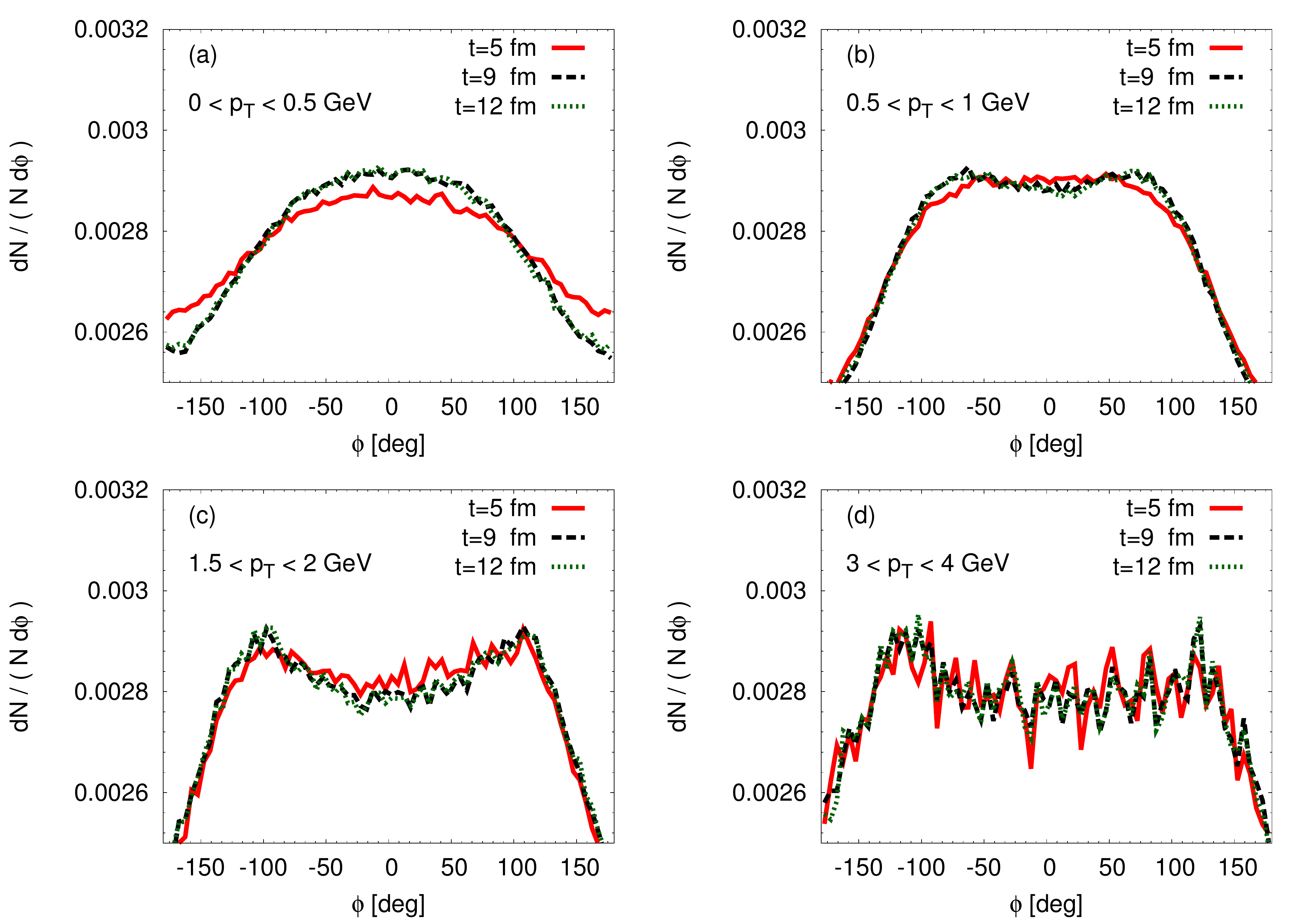}
\caption[Two-particle correlations for several $p_T$ regions extracted
from BAMPS calculations for jets in HIC for a fixed angle position
of $\phi_{\rm jet} = 180^{\circ}$ on the semi-circle]
{(Color online) Two-particle correlations, ${\rm d}N/(N{\rm d}\phi)$, extracted from BAMPS calculations
for a fixed value of $\eta/s = 0.08$ and various evolution times.
The jet is initialized at a fixed-angle position of $\phi_{\rm jet} = 180^{\circ}$ on the semi-circle (see Fig.~\ref{fig:numMach:hic_jetB_evolution_Fixed180}).}
\label{fig:numMach:hic_TPC_jetB_eCuts_Fixed180_5GeV_variousTimes}
\end{center}
\end{figure}

Please note that the local energy density in Fig.~\ref{fig:numMach:hic_jetB_evolution_Fixed180}
at $t = 12$ fm/$c$ is much smaller than the
typically expected $e_{\rm crit} \approx 0.6$ $ \rm GeV/fm^3$ in the deconfined phase \cite{Xu:2008av}. We mentioned above that BAMPS has no effective hadronization procedure;
thus, we extract all observables from the final gluon distribution. However, Fig.~\ref{fig:numMach:hic_TPC_jetB_eCuts_Fixed180_5GeV_variousTimes} proves that
the extracted azimuthal two-particle correlations do not change significantly after $t= 5-6$ fm/$c$.
While the
cross section increases to unphysical values for energy densities smaller than the
critical energy density $e_{\rm crit} \approx 0.6$ $\rm GeV/fm^3$, those energy
densities are thus so small that the interaction rate of the system decreases rapidly.
This implies that the further evolution of the deconfined phase
with the same $\eta/s$ after $t= 5-6$ fm/$c$ is reasonable. 
%
%

\subsection{Scenario II}

\begin{figure*}[tp!]
 \begin{center}
 \includegraphics[width=\textwidth]{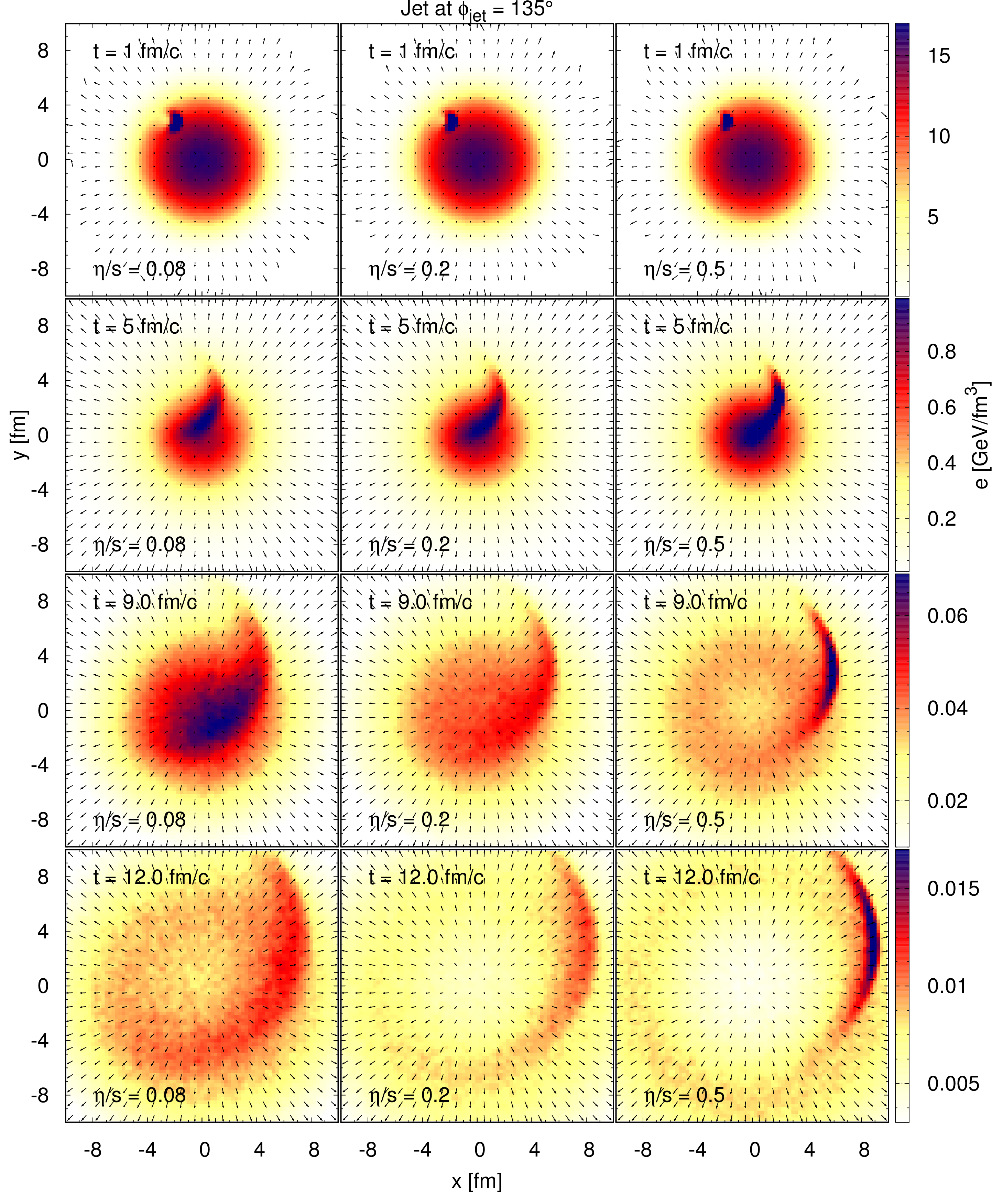}
 \caption[Time evolution of Mach cones in a central heavy-ion collision with a jet
 originating from a fixed angle position of $\phi_{\rm jet} = 135^{\circ}$ on the semi-circle]
 {(Color online) Time evolution of a Mach cone in a central heavy-ion collision. We show the local-rest frame 
 energy density within a mid-rapidity range of $\left| \eta_{\rm rap} \right| < 0.1$,
 overlaid by the velocity profile indicated by scaled arrows.
 The results are depicted at various time steps and for
 different values of the ratio of shear viscosity over entropy density, $\eta/s$.
 The jet is initialized at a fixed-angle position of $\phi_{\rm jet} = 135^{\circ}$ on the semi-circle
 with an initial momentum of $p_x = E_{\rm jet} = 20$ GeV.}
 \label{fig:numMach:hic_jetB_evolution_Fixed135}
 \end{center}
 \end{figure*}

In a second scenario, we consider a single jet with a fixed-angle
position of $\phi_{\rm jet} = 135^{\circ}$ on the semi-circle 
to study the effects of deflection on the jet due to radial flow.
The time evolution for different values of the ratio of shear viscosity over entropy density,
$\eta/s$ is depicted in Fig.~\ref{fig:numMach:hic_jetB_evolution_Fixed135}.
Due to the relative position on the semi-circle and the strong radial flow,
the single jet is deflected, especially for small $\eta/s$.
This deflection, however, reduces for large $\eta/s$ as
the jet-medium interaction weakens, keeping the jet on its initial propagation direction.

A direct comparison to \textbf{scenario I} indicates
that the jet traverses a significantly lower energy-density region, thus 
reaching only much lower values for the energy density in the shock-front region, especially 
for $\eta/s = 0.08$. For lower energy-density regions, however, the mean free path of the jet 
is larger and results in a reduced jet quenching. However, 
a medium with $\eta/s = 0.08$ is characterized by a strong collective behavior. 
The radial flow leads to a strongly distorted Mach cone pattern which is 
significantly reduced in a medium with $\eta/s = 0.5$.

%
\begin{figure}[ht!]
\begin{center}
\includegraphics[width=\columnwidth]{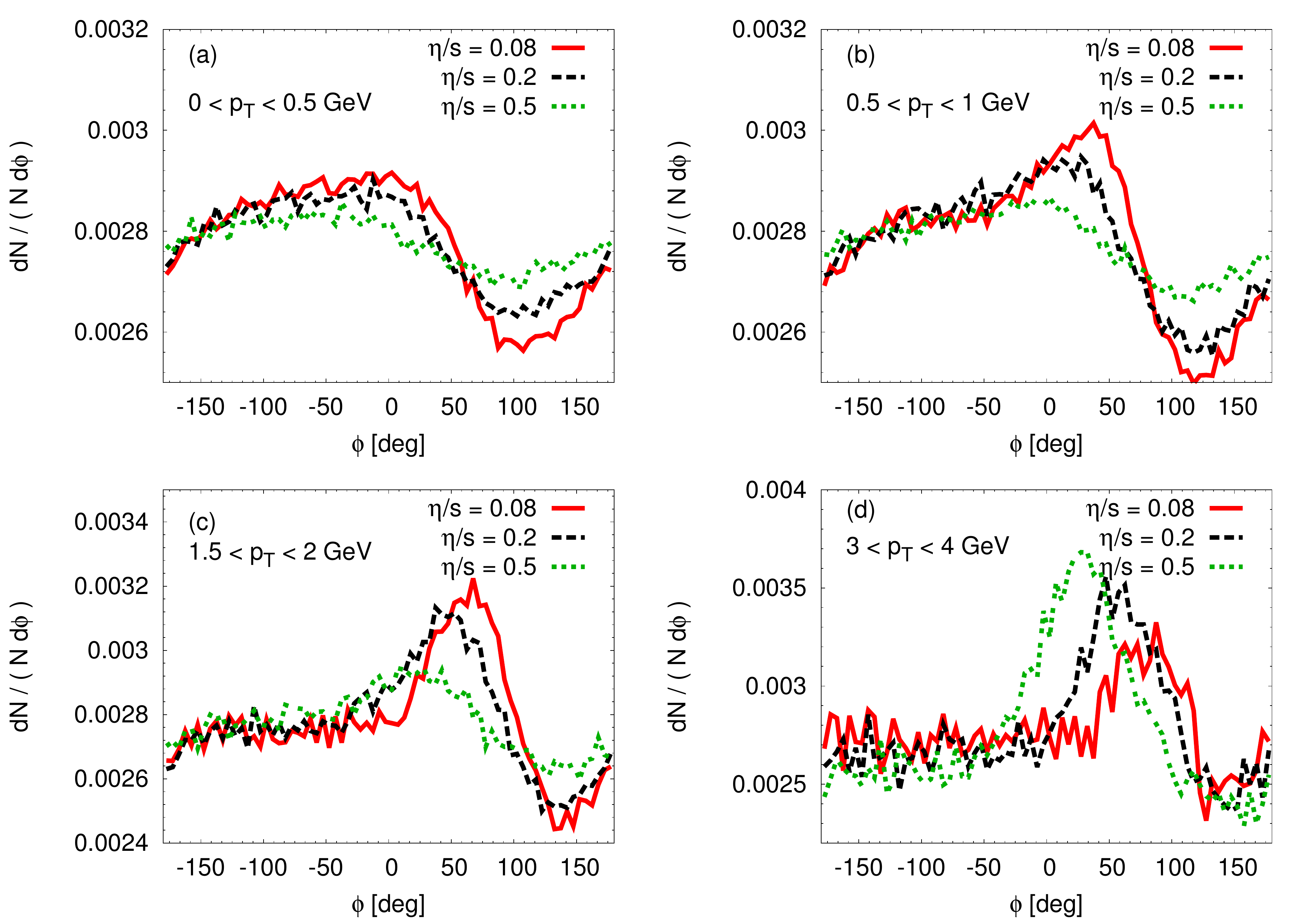}
\caption[Two-particle correlations for several $p_T$ regions extracted
from BAMPS calculations for jets originating from a fixed angle position
of $\phi_{\rm jet} = 135^{\circ}$ on the semi-circle.]
{(Color online) Two-particle correlations, ${\rm d}N/(N{\rm d}\phi)$, extracted from BAMPS calculations
for different values of $\eta/s$ and $p_T$-cuts.
The jet is initialized at a fixed-angle position of $\phi_{\rm jet} = 135^{\circ}$ on the semi-circle (see Fig.~\ref{fig:numMach:hic_jetB_evolution_Fixed135}).
The results are shown for a fixed time $t=12$ fm/$c$.}
\label{fig:numMach:hic_TPC_jetB_eCuts_Fixed135_5GeV}
\end{center}
\end{figure}
%

These results indicate that the generated diffusion wake
and head shock are less compensated by the radial flow in \textbf{scenario II}
due to the fact that the jet does not propagate in the direction opposite to the radial
flow. The deflection of the jet for
$\eta/s = 0.08$ leads to only one peak in the two-particle correlations
as displayed in Fig.~\ref{fig:numMach:hic_TPC_jetB_eCuts_Fixed135_5GeV}.
This peak originates from the head shock and diffusion wake of the distorted Mach cone and becomes
sharper and more pronounced for larger $p_T$.
We also see that the position of the induced peaks in 
Fig.~\ref{fig:numMach:hic_TPC_jetB_eCuts_Fixed135_5GeV} changes with viscosity. 
The larger the value of $\eta/s$, the smaller the peak angle, reflecting the fact that 
the jet is less deflected for larger values of $\eta/s$.

Although the two-particle correlations determined for this specific single-jet scenario
only exhibit one single peak, a double-peak structure is possible
when considering a second jet event in the lower semi-circle \cite{Betz:2010qh,Renk:2006mv,Chaudhuri:2006qk,Chaudhuri:2007vc}.
In this case, however, the double peak is mainly generated due to the
superposition of several deflected and distorted jet-induced Mach cones;
i.e., the origin for the double-peak structure is not the Mach cone itself 
but the deflected head shocks and diffusion wakes of the distorted Mach cones
as demonstrated below in \textbf{scenario III}.

\subsection{Scenario III}

\begin{figure}[t!]
\begin{center}
\includegraphics[width=\columnwidth]{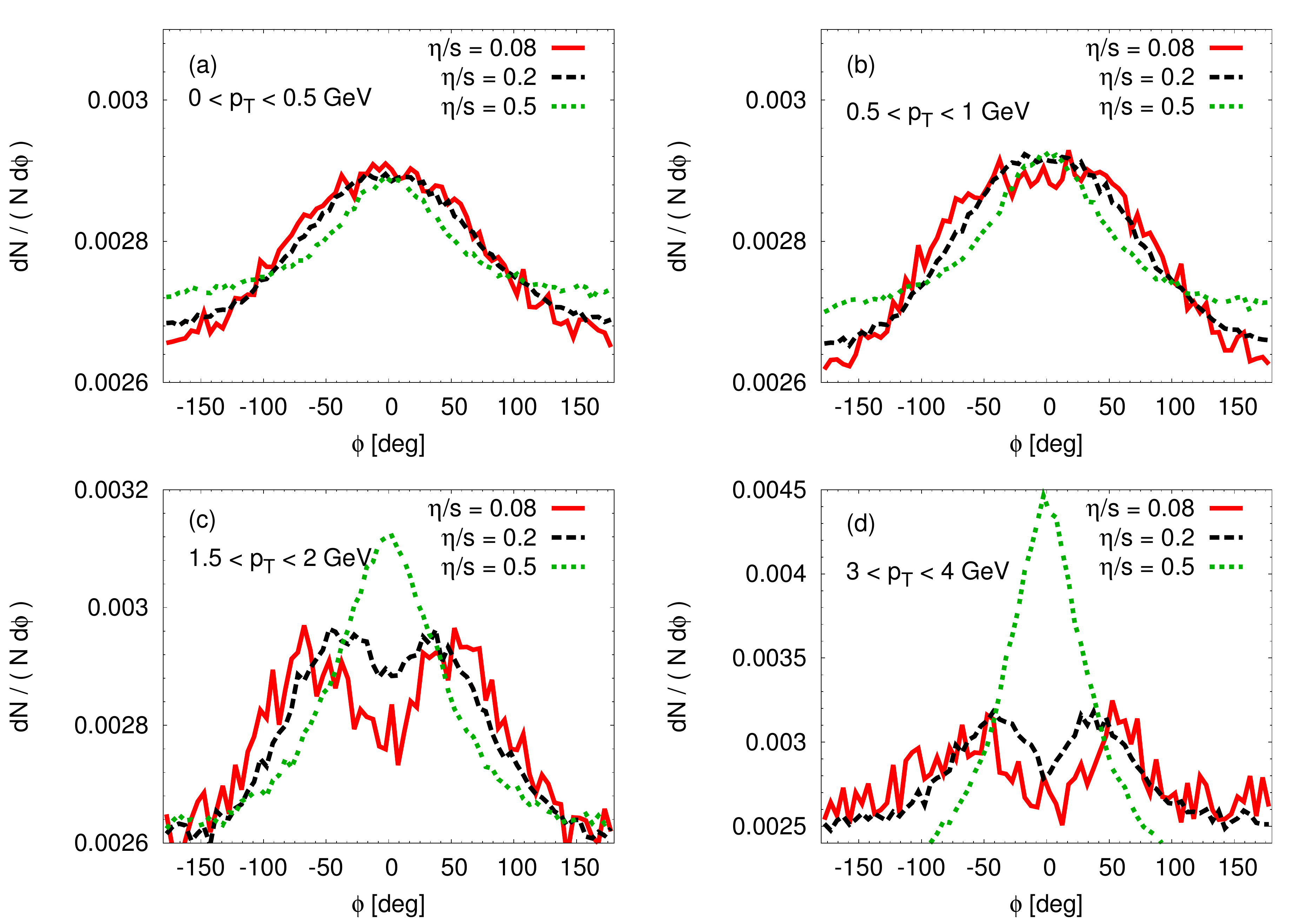}
\caption[Two-particle correlations for several $p_T$ regions extracted
from BAMPS calculations for jets in HIC with randomly starting position]
{(Color online) Two-particle correlations, ${\rm d}N/(N{\rm d}\phi)$, extracted from BAMPS calculations
for different values of $\eta/s$ and $p_T$-cuts. Here, we consider various starting
positions of the jet on the semi-circle. The results
are shown at a fixed time of $t=12$ fm/$c$.}
\label{fig:numMach:hic_TPC_jetB_eCuts_5GeV}
\end{center}
\end{figure}

Averaging over all possible jet trajectories with randomly
chosen starting positions on the semi-circle, i.e., $\phi_{\rm jet} =90-270^{\circ}$,
\textbf{scenario III} gets closest to the experimental situation.
Here, in contrast to \textbf{scenarios I} and \textbf{II}, 
we consider many different jet events. Figure~\ref{fig:numMach:hic_TPC_jetB_eCuts_5GeV} 
displays the normalized azimuthal particle distribution determined for this
multiple-jet scenario. A double-peak structure appears for sufficiently high $p_T$ and
low $\eta/s$ with peaks around $\phi \approx 50^{\circ}$. This result is in line 
with Ref.\ \cite{Betz:2010qh} applying a hydrodynamic evolution.

%

\section{Discussion}

The above results provide evidence that Mach cones can form in a heavy-ion collision if
the medium created has a small viscosity. However, as compared to an ideal, zero-viscosity medium, 
the pattern of the Mach cone is more curved. If a jet gets deflected by 
radial flow, the shape of the induced Mach cone is distorted.
We demonstrate that the radial flow in a heavy-ion collision affects both, the
final pattern of the Mach cone \cite{Satarov:2005mv} and the observed 
particle distribution. 

Considering a single jet event for a sufficiently
small value of $\eta/s = 0.08$, the interplay of jet-induced Mach cones
and radial flow may lead to a double-peak structure,
as shown in \textbf{scenario I}.
In this scenario the double-peak
structure originates from the Mach cone since the contributions of head shock
and diffusion wake are canceled by the radial flow.

For a single jet not traversing the medium in direction opposite to
the radial flow, however, the double-peak structure does not appear;
cf.\ \textbf{scenario II}. Here, the radial flow leads to a 
deflection of the jet, resulting in a single peak arising in the
two-particle correlations. Nonetheless, considering a multiple-jet scenario as in \textbf{scenario III},
a double-peak structure occurs as well that is mainly generated
by the superposition of the deflected and distorted head shocks
and diffusion wakes of the jet-induced Mach cones \cite{Betz:2010qh,Renk:2006mv}.

We demonstrate above that a sufficiently large value
of $\eta/s$ tends to destroy the characteristic
structure of a Mach cone in a three-dimensional expanding heavy-ion collision
and results in a single peak in the
two-particle correlation. 
Here, the contributions from both head shock and diffusion wake 
are less canceled by the radial flow than in the case of a smaller value of $\eta/s$. 
Moreover, larger dissipation reduces the deflection of jets. 
However, not only viscosity tends to destroy
the double-peak structure, but also the strength
of the jet. A jet with a large initial
energy punches through the medium and the effect
of the radial flow weakens. For such a high-energy jet, 
the diffusion wake and head-shock contribution in \textbf{scenario I}
would overshadow the double-peak structure, while
in \textbf{scenario III} most of the jets would get less
deflected. However, a jet with a very small initial energy density
would not lead to a double-peak structure either, as in this case the strong radial
flow observed in a heavy-ion collision does not allow a Mach cone 
to be developed.

Our results imply that the influence of the radial
flow is most important for the observation
of a double-peak structure. A large dissipation and/or a large initial jet energy
leads to a smaller influence of the radial flow.
The effect of longitudinal flow is studied here in a
qualitative way and was not considered in previous studies \cite{Betz:2010qh}. 
We conclude that the longitudinal flow does not affect the
results significantly.

The Mach-cone angle extracted from two-particle
correlations is certainly not connected to the equation-of state \cite{Satarov:2005mv}.
\textbf{Scenario III} provides evidence that
the contribution of this double-peak structure
mainly originates from the head shocks and diffusion wakes of the jet-induced Mach cones.
To extract the Mach angle and thus determine the speed of sound of the medium
would require the measurement of a single jet event as shown
in \textbf{scenario I} which remains an extremely tough experimental challenge.

\section{Conclusion}

In conclusion, we have shown that Mach cones can form in a longitudinal and
transverse expanding heavy-ion collision. Considering multiple-jet events, 
a double-peak structure evolves if the ratio of shear viscosity over entropy density
is $\eta/s<0.5$. Please note that the peak position does not change 
significantly with $\eta/s$. However, this double-peak structure does not
originate from Mach cones but arises due to the contribution of the head shock
and diffusion wake originating from the deflected jets \cite{Satarov:2005mv,Betz:2010qh}. 

Although initial-state fluctuations certainly provide a 
large contribution to the experimentally observed double-peak structure \cite{Takahashi:2009na,Schenke:2010rr,Bhalerao:2011bp,Ma:2010dv,Ayala:2012bv,Aamodt:2011by,ATLAS:2012at,Chatrchyan:2012wg},
it will be important to quantify the contribution from jets. The double-peak
signal might not be an appropriate observable to disentangle both contributions 
\cite{Bouras:2012mh} as it is mainly medium driven. However, allocating a more appropriate
signature remains an open challenge.

%
 \section*{Acknowledgements}
%
The authors are grateful to H.\ Niemi, E. Molnar, D.\ H.\ Rischke, J.\ Noronha,
G. \ Torrieri, P. \ Huovinen, J. Ulery, and H.\ St\"ocker
for discussions and to the Center for Scientific Computing (CSC)
at Frankfurt University for providing the necessary computing resources. 
I.B. acknowledges support from HGS-Hire. Z.X. work is supported
partially by the Major State Basic Research Development Program in
China (Grant No. 2014CB845400) and by the National Natural Science Foundation
of China projects (Grants No. 11275103, 11335005). This work was supported by BMBF
and the Helmholtz International Center for FAIR within the framework of the LOEWE
program launched by the State of Hesse.

\end{document}